\documentclass[
    ,final            
  ]
  {aipproc}

\layoutstyle{8x11double}

\begin{document}

\title{A Search for Pulsed and Bursty Radio Emission from X-ray Dim Isolated Neutron Stars}


\author{V.I.~Kondratiev}{
  address={West Viriginia University, Physics Department, Morgantown, USA}
  ,altaddress={National Radio Astronomy Observatory, USA}
}

\author{M.~Burgay}{
  address={INAF, Osservatorio Astronomico di Cagliari, Italy}
}

\author{A.~Possenti}{
  address={INAF, Osservatorio Astronomico di Cagliari, Italy}
}

\author{M.A.~McLaughlin}{
  address={West Viriginia University, Physics Department, Morgantown, USA}
}

\author{D.R.~Lorimer}{
  address={West Viriginia University, Physics Department, Morgantown, USA}
}

\author{R.~Turolla}{
  address={Universita\textquotesingle ~di Padova, Italy}
}

\author{S.~Popov}{
  address={Sternberg Astronomical Institute, Moscow, Russia}
}

\author{S.~Zane}{
  address={Mullard Space Science Laboratory, London, UK}
}

\begin{abstract}
We have carried out a search for radio emission from six X-ray dim isolated neutron stars (XDINSs) 
observed with the Robert C. Byrd Green Bank Radio Telescope (GBT) at 820~MHz. No bursty or pulsed 
radio emission was found down to a $4\sigma$ significance level. The corresponding flux limit is 0.01--0.04~mJy depending 
on the integration time for the particular source and pulse duty cycle of 2\%. These are the most sensitive limits yet
on radio emission from these objects.
\end{abstract}

\maketitle

\section{Introduction}
The \emph{ROSAT} mission discovered a group of seven nearby low-luminosity isolated neutron stars, termed 
the X-ray dim isolated neutron stars (XDINSs).
They share very similar properties and are characterized by soft blackbody-like spectra in the 
range $\sim 40$--100 eV, very faint optical counterparts ($V>25$), long spin periods of 3--12~s (see Table~\ref{table}).
For a recent review see \citet{haberl2007}.

So far, no confident detections of pulsed radio emission were found from XDINSs. The aim of this project is to 
search for pulsed and bursty radio emission from XDINSs to link them in their evolutionary 
scenarios with other classes of neutron stars, such as magnetars and rotating radio transients (RRATs). 
All three populations of neutron stars have similar properties, such as period, period derivative, age, 
and magnetic field so that connections between them are plausible.

\section{Observations and Data Processing}
The observations were conducted with the GBT on May 28--31, 2006 at a center frequency of 820~MHz. The Spigot pulsar
backend was used to record a signal with $81.92~\mu$s time resolution in 1024 channels covering a 50-MHz receiver bandpass.

All known XDINSs except for RX~J0420.0$-$5022 were observed. Table~\ref{table} lists the name of the source,
X-ray period P, the total observing time T${}_\mathrm{obs}$, 
the largest trial value DM${}_\mathrm{top}$ of dispersion measure (DM), and the
$4\sigma$ upper limit on  pulsed radio emission, S${}_\mathrm{lim}$, for pulse duty cycle of 2\%.
The chosen DM range was determined by the NE2001 model of 
free electron density in our Galaxy~\citep{ne2001}. We assumed distances $< 1$~kpc for all sources
but J1856.5$-$3754 and minimum pulse widths of 1~ms. For J1856.5$-$3754 the DM${}_\mathrm{top}$ corresponds
to the distance of 250~pc. This is half as much again as the true value of $161^{+18}_{-14}$~pc based on the parallax
measurement by \citet{kaplan2007a}. For RX~J0720.4$-$3125, the chosen distance of $< 1$~kpc is also compatible with
doubled value of parallax distance of $360^{+170}_{-90}$~pc \citep{kaplan2007b}.

\begin{table}[tbhp]
\begin{tabular}{lcccc}
\hline
\tablehead{1}{l}{c}{XDINS}
  & \tablehead{1}{c}{c}{P\\(s)}
      & \tablehead{1}{c}{c}{$\mathbf{T}_\mathbf{obs}$\\(h)}
        & \tablehead{1}{c}{c}{DM${}_\mathbf{top}$ \\(pc cm${}^\mathbf{-3}$)}
	  & \tablehead{1}{c}{c}{S${}_\mathbf{lim}$\\($\mu$Jy)} \\
\hline
J0720.4$-$3125 & 8.39 & 4 & 98 & 10 \\
J0806.4$-$4123 & 11.37 & 2 & 172  & 40 \\
J1308.6+2127 & 10.31 & 4.25 & 27 & 20 \\
J1605.3+3249 & 6.88 & 4 & 50 & 10 \\
J1856.5$-$3754 & 7.055 & 4 & 12 & 20 \\
J2143.7+0654 & 9.44 & 4.08 & 33 & 20 \\
\hline
\end{tabular}
\caption{Parameters and processing details for six observed XDINSs.}
\label{table}
\end{table}

The data were processed using both \emph{SIGPROC}\footnote{\url{http://sigproc.sourceforge.net}}
and \emph{PRESTO}\footnote{\url{http://www.cv.nrao.edu/~sransom/presto}} packages.
First, the data were filterbanked into the ``Sigproc'' format and decimated by 6 samples.
After an excision of radio frequency interference (RFI),  
the data were dedispersed for a number of
different DMs from zero to DM${}_\mathrm{top}$ (see Table~\ref{table}) with a step size of
roughly 1~pc cm${}^{-3}$. Finally, a single-pulse search, FFT search, and Fast Folding 
Algorithm (FFA) search were applied for every dedispersed time series. A new FFA search
code \emph{ffasearch} was written to implement the FFA 
algorithm~\citep{staelin1969}.
This code has been successfully proven to detect known long-period pulsars with higher 
signal-to-noise ratios (SNR) than traditional FFT searches.

\section{Single-Pulse Search}
The single-pulse search is a very powerful tool to detect pulsating sources of strong individual 
bursts that are too weak to be detected through a regular FFT search. The RRATs were discovered 
in this way~\citep{mmclaugh2006}. XDINSs show possible connection with the RRATs, thus making 
the single-pulse search very important.

\begin{figure}
 \includegraphics[angle=270,scale=0.31]{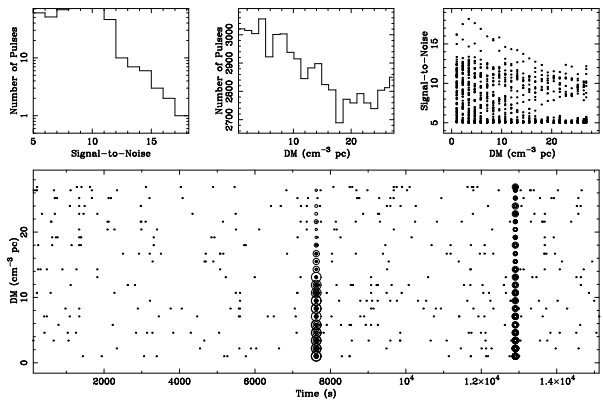}
 \caption{The single-pulse search diagnostic plot for 1RXS~J1308.6+2127. {\it Top:} the histograms of
 the number of bursts versus SNR and DM, and the dependence
 of SNR versus DM {\it (from left to right)}. {\it Bottom:} the main time-DM plot with
 points' size corresponding to SNR of the burst. The detection threshold is $5\sigma$.
 Two vertical lines of bursts on the bottom plot are either stronger at zero DM or pretty constant
 over all the DM range. Both are RFI signatures rather than real pulse which would appear as 
 vertical segment of bursts centered at its strongest burst at non-zero DM with SNR of other 
 bursts gradually fading to the edges of the segment. If there is a train
 of such segments at the same non-zero DM than it is surely a manifestation of burst emission
 of the pulsar.
 }
 \label{sp}
\end{figure}

To detect spiky emission from XDINSs, we have performed the single-pulse search for each isolated
neutron star in our analysis. An example diagnostic plot for the single-pulse search for
1RXS~J1308.6+2127 is shown in Figure~\ref{sp}. The distinctive feature of a candidate source
is the train of strong pulses at some non-zero DM.

Unfortunately, almost all our data were seriously contaminated by RFI 
of external and equipment nature. The vertical lines of bursts on the bottom plot of Fig.~\ref{sp} 
are RFI signatures that were 
still in the data after RFI excision. The powerful RFI masking program \emph{rfifind} from \emph{PRESTO} 
package turned out to be inefficient, because it does not deal with broad impulse non-periodic RFI. 
Thus, we used a dedicated program \emph{rfimarker} for all sources except for RX~J1605.3+3249.

The search was done for a range of pulse 
widths using matched filtering techniques~\citep{mmclaugh2003}. Only the candidate with largest 
SNR was plotted. No apparent strong pulses were found. This could be due to contaminated data, 
or a lack of bursty emission from the studied XDINSs, or a rate of burst occurrence of less 
than 1 every few hours.

\section{FFT and FFA Searches}
Though the FFT search is not efficient for long period sources, because of low-frequency 
noise, we still applied it to every observed XDINS. However, as we expected we did not find 
any promising candidates.

\begin{figure}
 \includegraphics[angle=270,scale=0.28]{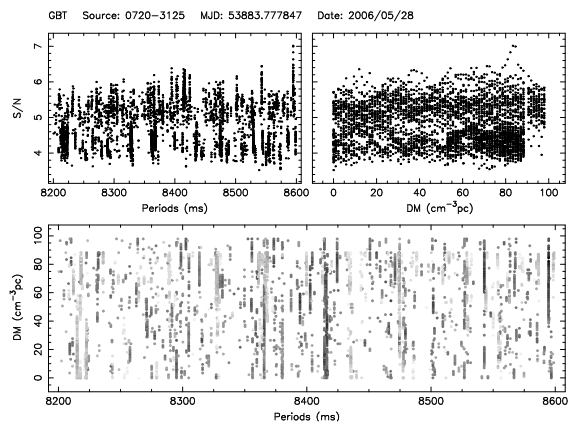}
 \caption{Example of an FFA diagnostic plot for RX~J0720.4$-$3125. The {\it top-left} plot is the 
 periodogram, i.e. folded profile's significance for every period candidate, the {\it top-right} plot shows
 the profile significance versus DM. The plot on the {\it bottom} represents the SNR of folded
 profiles versus DM and the period. The darker points have larger SNRs. Signature of the pulsar on this
 plot is similar to that for a single-pulse search. Long lines over broad DM range are not real but
 rather a strong RFI of periodic origin.
 }
 \label{ffa}
\end{figure}

On the contrary, the FFA search should be very effective. We performed the FFA search for 
period ranges within a few hundred milliseconds around the actual X-ray period. Figure~\ref{ffa} shows 
an example of an FFA diagnostic 
plot for RX~J0720.4$-$3125.  It is similar to that for the single-pulse search (see Fig.~\ref{sp}). 

The presence of the a real pulsed source would show up as elongated train of candidates in both period 
and DM axes. We have inspected the folded profiles for many candidates and did not find any 
significant profiles down to the $4\sigma$ level, corresponding to 0.01--0.04~mJy depending on the 
integration time for the particular source and duty cycle of 2\%.

\section{Summary}
\begin{figure}
 \includegraphics[scale=0.44]{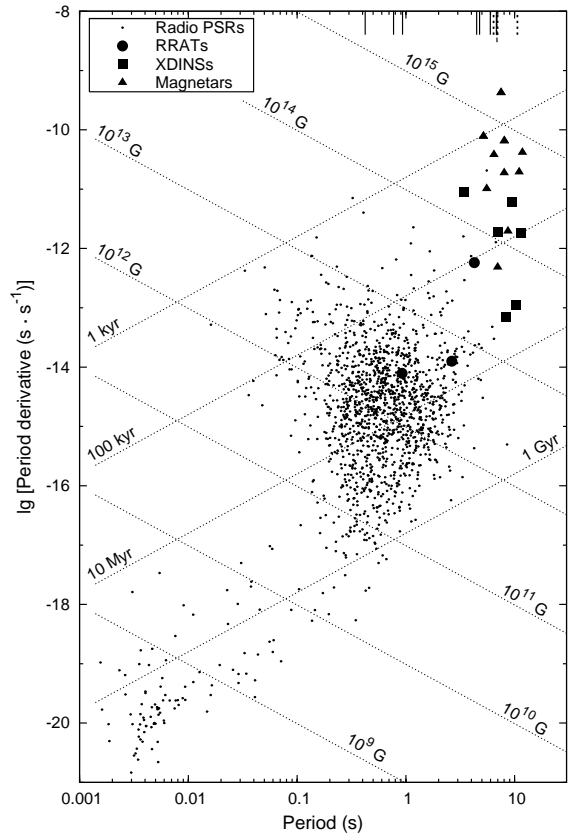}
 \caption{P-\.P diagram. Lines on the top 
mark the period values for the objects with unknown yet period derivatives: seven 
RRATs (solid), one magnetar SGR~1627$-$41 and two candidates AX~J1845$-$03 and
CXO~J164710.2$-$455216 (dashed), and one XDINS RX~J1605.3+3249 (longer dashed).
 }
 \label{ppdot}
\end{figure}

The sporadicity of the RRATs' radio emission led to immediate suggestions that they are 
related to other classes of traditionally ``radio-quiet'' neutron stars such as XDINSs and 
magnetars. \citet{popov2006} have shown that the implied birthrate of 
RRATs is more  consistent with that of XDINSs than that of magnetars. As shown in the P-\.P 
diagram on the Figure~\ref{ppdot}, RRATs and XDINSs also have similar periods and period 
derivatives, implied ages and magnetic fields. However, the RRATs spin-down properties are 
also consistent with those of the normal pulsar population and X-ray observations of one 
RRAT~\citep{mmclaugh2007} reveal properties similar to those of both normal radio pulsars
and XDINSs.

The RRATs are powerful sources of isolated radio bursts, and we have not detected such bursts, 
or any periodic emission, from the six XDINSs we have observed. Because 
the distances to the XDINSs are believed to be much smaller than those to the RRATs, we should 
have had high sensitivity to RRAT-like radio emission. However, our non-detection of such emission 
does not necessarily mean that there is no relationship between these two source classes.

XDINSs may simply be ``radio-quiet'', but it is also likely that perhaps the narrow radio beams 
from these XDINSs are simply misaligned with our line-of-sight. It is possible that searches at 
lower frequencies, where radio emission beams are believed to be wider, may be more sensitive to 
radio emission from XDINSs. Indeed, Malofeev and co-authors~\citep{malofeev2005, malofeev2007} 
reported detection of radio emission from RX~J1308.6+2127 and  RX~J2143.7+0654 at the low frequency 
of 111~MHz. On the other hand, XDINSs could have very steep spectral indices. If the detection 
of Malofeev and co-authors is real, our non-detection of radio emission from these two XDINSs 
at 820~MHz sets a lower limit on the spectral index of 3.6. Finally, it is possible that our 
non-detection of radio emission from these XDINSs is due to the large amount of contamination 
from RFI. Our search highlights the importance of improved excision algorithms for impulsive, 
broadband terrestrial interference.


\begin{theacknowledgments}
SZ thanks STFC (ex-PPARC) for support through an AF.
The Robert C. Byrd Green Bank Telescope (GBT) is
operated by the National Radio Astronomy Observatory which is a
facility of the U.S. National Science Foundation operated under
cooperative agreement by Associated Universities, Inc.
\end{theacknowledgments}


\bibliographystyle{aipproc}   


\bibliography{xdins}


\end{document}